\PassOptionsToPackage{table}{xcolor}
\documentclass[format=acmsmall, review=false, screen=true, anonymous=false]{acmart}

\usepackage{booktabs} 
\usepackage[ruled]{algorithm2e} 
\usepackage{natbib}
\usepackage{enumitem}
\usepackage{tabularx}
\usepackage{siunitx}
\usepackage{multirow}
\usepackage{subcaption}
\usepackage[justification=centering]{caption}
\usepackage{titlesec}

\newenvironment{itquote}
{\begin{quote}\itshape}
{\end{quote}}

\definecolor{lavenderblue}{rgb}{0.9, 0.9, 0.98}

\acmJournal{PACMHCI}
\acmVolume{3}
\acmNumber{CSCW}
\acmArticle{55}
\acmYear{2019}
\acmMonth{11}
\acmPrice{15.00}
\acmDOI{10.1145/3359157}

\setcopyright{acmlicensed}

\received{April 2019} 
\received[revised]{June 2019}
\received[accepted]{August 2019}

\begin{document}
\title[Moderation Challenges in Voice-based Online Communities on Discord]{Moderation Challenges in Voice-based Online Communities on Discord}

\author{Jialun ``Aaron'' Jiang}
\affiliation{
  \institution{University of Colorado Boulder}
  \department{Department of Information Science}
  \streetaddress{ENVD 201, 1060 18th St.}
  \city{Boulder}
  \state{CO}
  \postcode{80309}
  \country{USA}}
\email{aaron.jiang@colorado.edu}

\author{Charles Kiene}
 \orcid{0000-0002-7327-8391}
 \affiliation{%
  \institution{University of Washington}
  \department{Department of Communication}
  \city{Seattle}
  \state{WA}
  \postcode{98195}
  \country{USA}}
\email{ckiene@uw.edu}

\author{Skyler Middler}
\affiliation{
  \institution{University of Colorado Boulder}
  \department{Department of Information Science}
  \streetaddress{ENVD 201, 1060 18th St.}
  \city{Boulder}
  \state{CO}
  \postcode{80309}
  \country{USA}}
\email{skyler.middler@colorado.edu}

\author{Jed R. Brubaker}
\affiliation{
  \institution{University of Colorado Boulder}
  \department{Department of Information Science}
  \streetaddress{ENVD 201, 1060 18th St.}
  \city{Boulder}
  \state{CO}
  \postcode{80309}
  \country{USA}}
\email{jed.brubaker@colorado.edu}

\author{Casey Fiesler}
\affiliation{
  \institution{University of Colorado Boulder}
  \department{Department of Information Science}
  \streetaddress{ENVD 201, 1060 18th St.}
  \city{Boulder}
  \state{CO}
  \postcode{80309}
  \country{USA}}
\email{casey.fiesler@colorado.edu}

\renewcommand{\shortauthors}{J. A. Jiang et al.}

\begin{abstract}
Online community moderators are on the front lines of combating problems like hate speech and harassment, but new modes of interaction can introduce unexpected challenges. In this paper, we consider moderation practices and challenges in the context of real-time, voice-based communication through 25 in-depth interviews with moderators on Discord. Our findings suggest that the affordances of voice-based online communities change what it means to moderate content and interactions. Not only are there new ways to break rules that moderators of text-based communities find unfamiliar, such as disruptive noise and voice raiding, but acquiring evidence of rule-breaking behaviors is also more difficult due to the ephemerality of real-time voice. While moderators have developed new moderation strategies, these strategies are limited and often based on hearsay and first impressions, resulting in problems ranging from unsuccessful moderation to false accusations. Based on these findings, we discuss how voice communication complicates current understandings and assumptions about moderation, and outline ways that platform designers and administrators can design technology to facilitate moderation.
\end{abstract}

%
%


%
%



\maketitle

\section{Introduction}

Online communities face malicious behaviors like hate speech \cite{chandrasekharan_you_2017} and harassment \cite{blackwell_classification_2017}, and many community moderators volunteer their time to combat these problems on a daily basis. People in online communities usually have a good sense of how moderation works: Someone posts something inappropriate; either it is seen and removed by a human moderator, or instantly removed by an automated moderator. This process is straightforward and easy to understand, and has long been part of people's mental models. But how does a moderator, whether human or automated, moderate inappropriate speech when it is spoken, in a real-time voice chat rather than text that can be erased? 

Adoption of new technology and new communication media introduces new ways for community members to engage and communicate, but it also introduces new norms to the communities \cite{orlikowski_using_2000}, and consequently changes what it means to moderate content and interactions. Consider real-time voice chat. This is not a new technology, but only recently with the increasing popularity of Discord and other voice-based online communities, has become more relevant for the everyday process of content moderation. In traditional text-based communities, moderation work mostly involves moderators locating the problematic content, and then removing it and sometimes also punishing the poster. This is a process that many people would take for granted, but how does this process work in the context of real-time voice, a type of content that lasts for a short time without a persistent record? The moderation of ephemeral content raises a number of questions: How do moderators locate the content? How do moderators remove the content? How do moderators know who the speaker is? How do moderators know whether the rule breaking happened at all?

Voice has exposed new problems for moderation, particularly due to the absence of a persistent, written record as is common in other major large-scale moderated spaces.  With voice and its strict ephemerality making current content management options impossible, voice moderators have to face these questions every day, and develop their own tactics and workarounds. Though Discord is designed for voice communication, Discord users have appropriated the platform technology to also play other types of audio (e.g., music or even noise) have become a part of user interaction as well, and we will describe ways that this also plays into types of rule-breaking. Furthermore, these issues do not impact voice alone---answers to these questions will not only provide insights toward how voice moderation works, but also provide insights for a range of emerging types of technology where interactions and content are ephemeral, such as immersive online games (e.g., Fortnite) and virtual reality (VR). These insights will also inform design and policy for communities that adopt new technology in order to help mitigate potential moderation problems.

In this work, we investigate the work of moderators and the challenges they experience in real-time voice-based online communities on Discord, a popular voice-over-IP (VoIP) platform. Through an analysis of 25 interviews with volunteer moderators of Discord communities, we first describe new types of rules unique to voice and audio-based communities and new ways to break them. Then, we describe how moderators struggled to deal with these problems. Moderators tried to give warnings first but sometimes had to take actions based on hearsay and first impressions. To avoid making elaborate rules for every situation, moderators instead simply stated that they had highest authority. We then detail how these problems point to moderators' shared struggle---acquiring evidence of rule breaking, and how moderators' evidence gathering strategies could fail in different scenarios.

Through the lens of Grimmelman's taxonomy of community moderation \cite{grimmelmann_virtues_2015} that focuses on techniques and tools, we argue that voice precludes moderators from using the tools that are commonplace in text-based communities, and fundamentally changes current assumptions and understandings about moderation. From here, we provide recommendations for designing moderator tools for voice communities that automate much of moderators' work but still ultimately put humans in charge, and finally argue that platform administrators and moderators must consider the limitations imposed by the technological infrastructure before importing existing rules and moderation strategies into new communities.

\section{Related Work}

To situate our study, we begin by revisiting work on online community moderation, and voice as an example of new technology adopted in online communities. These two threads of research highlight how the unique characteristics of voice communication can not only bring new problems to moderation, but also exacerbate existing problems. Finally we provide a brief review of research on ephemerality, which we later show to be the source of the biggest challenge of moderating voice.

\subsection{Moderation and Its Challenges}

Effective regulation is one of the key factors that contributes to the success of an online community \cite{kraut_building_2014}. Though regulation can occur at multiple levels, including Terms of Service (TOS), many online communities have their own community-created rules that are enforced by volunteer moderators \cite{fiesler_reddit_2018}. Grimmelmann \cite{grimmelmann_virtues_2015} lists four basic techniques that moderators can use to moderate their communities: (1) excluding---to remove a member from the community; (2) pricing---to impose monetary costs on participation; (3) organizing---to delete, edit, or annotate existing content; and (4) norm-setting---to create desired, shared norms among community members. These techniques also differ in terms of the ways they are used (e.g., automatically or manually, or proactively or reactively). These techniques have become increasingly present in social computing and HCI scholarship, focusing on how they have been successful in addressing various problems in communities (e.g. \cite{chandrasekharan_you_2017, jhaver_online_2018, seering_moderator_2019}), ranging from using Reddit AutoModerator to automatically remove problematic content \cite{kiene_surviving_2016}, to setting positive examples to encourage similar behaviors in Twitch chat \cite{seering_shaping_2017}. 

While moderation is beneficial to communities, it is also challenging. Not only do moderators have to do the often thankless job of continuously looking at traumatic content and making personally uncomfortable decisions \cite{wohn_volunteer_2019, dosono_moderation_2019, roberts_commercial_2016, mcgillicuddy_controlling_2016}, they also often have to resort to imperfect solutions to problems in their communities. For example, while having clear, prominently displayed rules is helpful for community members to learn the norms, it may convey the message that these rules are often broken \cite{kraut_building_2014}, or make community members feel stifled and constrained \cite{kiene_surviving_2016}. The lack of clarity of the higher governing laws also made moderators' work difficult \cite{noauthor_content_nodate}. Furthermore, the limited availability of volunteer community moderators means that moderation is often delayed \cite{lampe_slashdot_2004}, leaving problematic content in place, and community members' efforts to circumvent moderation \cite{gerrard_beyond_2018, chancellor_thyghgapp:_2016} makes timely moderation even harder. Recognizing this tension, prior research has called for mutual understanding between community members and moderators \cite{jhaver_view_2018}. 

While there has been an emergence of machine-learning based automated moderation tools, it is difficult to gather enough training data for rule violations when rule breakers try to hide themselves, not to mention that these Automoderators may not be adaptive to new kinds of expression and communication \cite{gillespie_custodians_2018}. Even with the help of automated moderation tools, moderators still need to make nuanced, case-by-case punishment decisions that automated tools are not capable of \cite{seering_moderator_2019}. How to scale this human labor is still an open and challenging question with today's incredibly complex online communities with millions of users of diverse backgrounds, purposes, and values \cite{gillespie_custodians_2018}. Matias \cite{matias_civic_2019} calls out online moderation work as civic labor to recognize this constant negotiation of the meaning and boundary of moderation.

These challenges are widespread in many online communities, and new ones often appear in the context of new platforms and modes of interaction. For example, moderation practices on Twitch had to adapt to fast-paced, real-time content in ways that would not be necessary on a platform like Reddit \cite{wohn_volunteer_2019, seering_moderator_2019}. Our study continues to extend the current literature on moderation by examining these challenges (and new ones) in the context of voice chat.

\subsection{New Technologies and New Interactions}

The introduction of new technologies to social interactions often results in unexpected challenges in behavioral regulation--particularly as bad actors find new ways to break rules and push against community norms. For example, in Julian Dibbell's 1998 book \textit{My Tiny Life}, he describes ``a rape in cyberspace'' in which a user of the text-based community LambdaMOO used programmable avatars to control and assault other users \cite{dibbell_my_1998}.  As a result, community moderators had to change their rules and practices to deal with this unexpected use of the technology. New technology may result in new structures, as people enact norms through their continued use of it \cite{orlikowski_using_2000, shklovski_commodification_2009, lee_now_2011}. Though the technologies required to create programmable avatars were not new at the time they resulted in bad behavior in LambdaMOO, it may have been the first time that community moderators had to deal with how they might be used in that context. Similarly, voice communication technology has been around for more than 100 years, but it is a newer consideration for online community moderators.

\subsubsection{Voice Communication.}
Group voice communication goes back as far as ``party lines'' on telephone networks in the late 1800s. A party line was a shared telephone line for multiple households that had to be manually connected by human operators. There was no expectation of privacy on a party line because the operator as well as other people on the line could listen in at any time. As a result, AT\&T released guidelines around usage etiquette, encouraging people to cooperate. However, these rules were ``frequently broken,'' and eavesdropping, gossiping, and pranking were common problems \cite{noauthor_listening_1959, channel_at&t_nodate}. Though a very different context, we see analogs to the telephone company's struggle to moderate party lines in the struggles moderators shared in our study.

Prior research about voice communication in HCI and CSCW has focused on affordances of voice, often drawing comparison with video-based communication (e.g. \cite{oconaill_conversations_1993, olson_what_1995, isaacs_what_1994}). This line of research has revealed that video can increase the use of systems and results in better collaboration, but still suffers from problems like cumbersome turn-taking coordination, lack of peripheral cues, and difficult speaker determination (i.e., who ``holds the mic''). The roots of these problems suggest that they may be further exacerbated in voice-only communication due to the lack of visual cues, though that social norms may be helpful for mitigating some of the problems. For example, in evaluating real-time group voice communication as a social space, Ackerman et al. identified emerging norms, including explicit announcement of someone new joining or of unpredictable noises \cite{ackerman_hanging_1997}. Through five case-studies, Wadley et al. \cite{wadley_voice_2015} also showed that real-time voice was more susceptible to abuse, a finding that our study also resonates.

Recent research has examined voice communication in specific contexts. For example, Tang and Carpendale \cite{tang_mobile_2009} studied voice communication in a hospital, and identified the recurring problem of ambient noise in voice communication. Another thread of research looked at voice-based communities in rural India \cite{patel_avaaj_2010, vashistha_sangeet_2015}. Voice communication in these communities, however, is not real-time but instead based on playback of recorded audio snippets. This important characteristic not only made threaded conversations and actions such as rating and liking possible, but also enabled moderation tools that exist in text-based communities, such as deleting, categorizing, and ranking. More commonly, however, voice-based communication is real-time and ephemeral.

\subsubsection{Ephemerality.}

Ephemerality is a distinctive feature of real-time voice communication, as well as some contexts for text-based communication and online communities. Researchers have argued that forgetting is a crucial part of human experience \cite{mayer-schonberger_delete:_2009}. Ephemerality also facilitates spontaneous interactions between users, encourages experimenting with different personas, and reduces concerns about self-presentation \cite{bayer_sharing_2016, xu_automatic_2016, schlesinger_situated_2017}. Another line of research of anonymous communication points out the ephemerality of identity and how it allows people to explore the full range of their identity but subject people to the consequence of de-anonymization \cite{sharon_unpacking_2018}, raising the question of whether people can accurately estimate data persistence \cite{kang_strangers_2016}. The answer to this question, however, is that people tend to expect data persistence from platforms with default options being saving rather than deleting \cite{shein_ephemeral_2013}, and come up with saving strategies to deal with content, meaning, and context losses \cite{bernstein_4chan_nodate, cavalcanti_media_2017}. While these prior research showed people tried to save content only for personal consumption, persistent records are also critical to moderators' work as evidence of rule breaking. This study explores how gathering evidence becomes the biggest challenge in moderating the ephemeral voice.

\section{Research Site: Discord}
Discord\footnote{https://discordapp.com/} is a free cross-platform VoIP application that has over 200 million unique users as of December 2018. Communities on Discord are called ``servers,'' a term we will use throughout this paper to refer to these communities. Despite the term ``server,'' they are not self-hosted but instead hosted centrally on Discord hardware. While originally designed for video gaming communities as a third-party voice-chatting tool during gameplay, Discord servers now cover a wide range of topics such as technology, art, and entertainment. Every user can create their own servers as they wish, even simply as general chat rooms with no specific purpose. The size of Discord servers ranges from small groups of friends with a handful of people, to massive communities with hundreds of thousands of members. 

\begin{figure}
    \centering
    \includegraphics[scale=.3]{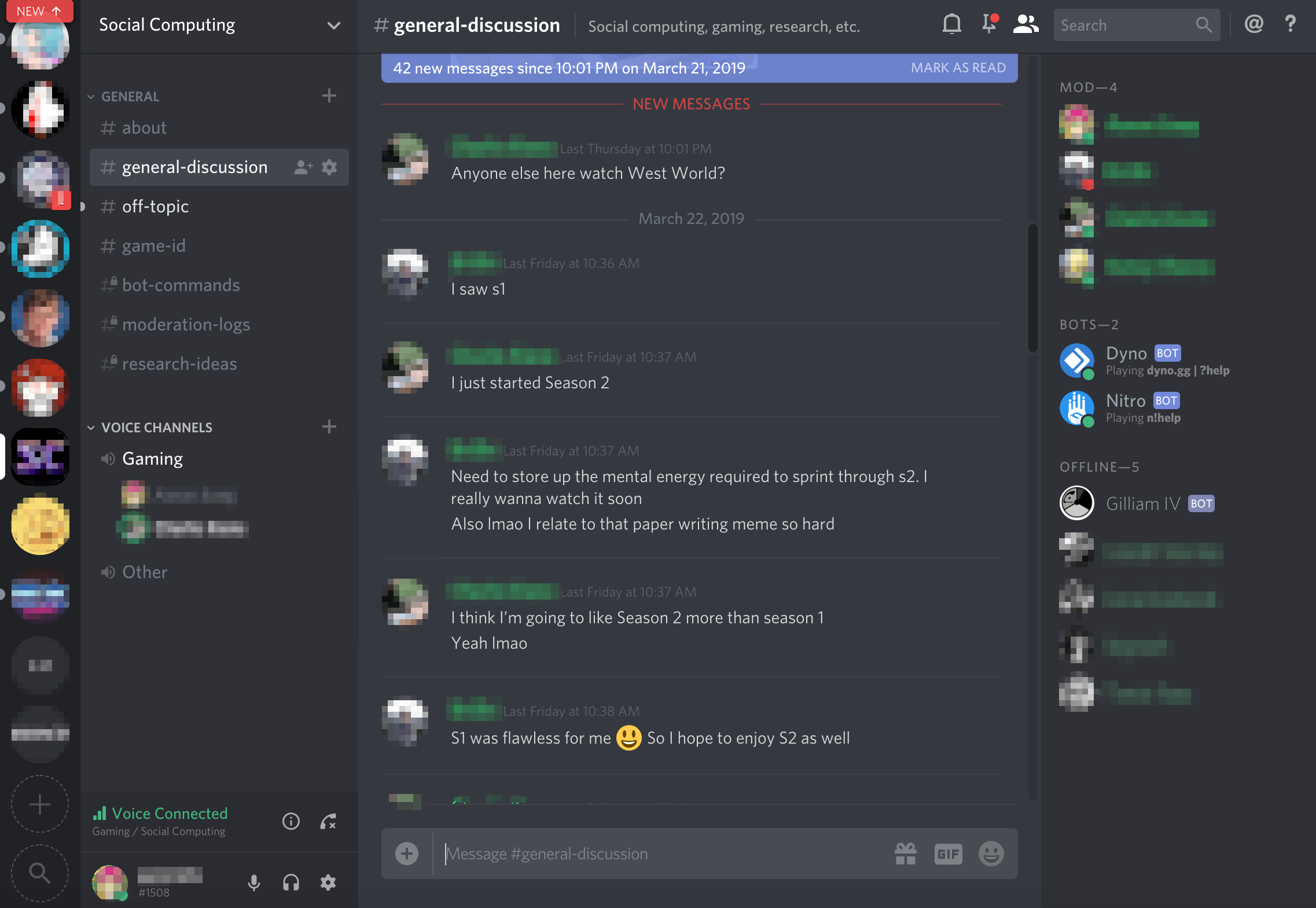}
    \caption{The Discord user interface. The far left sidebar lists all the Discord servers the user is a member of. The next side bar lists the text and voice channels of the Discord server the user is currently viewing. The middle area is for the scrolling text chat, and the right side bar lists the total users, categorized by their ``role.'' }
    \label{fig:vc}
\end{figure}

A server typically consists of separate virtual spaces called ``channels,'' usually with their own purposes, such as announcements or topic-specific conversations. A channel can be either a text channel or a voice channel, but not both. Users can also directly contact other users they are friends or share servers with through direct messages with text, voice, and video capabilities. A screenshot of the Discord interface is shown in Fig. \ref{fig:vc}.

In voice channels, the only means of communication is real-time voice chat. Users can choose to have their mic open all the time, or push a button to talk depending on their settings. Discord does not provide ways to record or store voice chat, making them ephemeral. Users currently in a voice channel will appear in the user list of the channel, and will disappear when they exit the channel. A green circle around a user's profile picture indicates the user is currently speaking. Users can also mute themselves---make themselves not be heard---or deafen themselves---make themselves not hear everyone else \emph{and} not be heard. Some Discord servers also have a special type of voice channel called ``music queue,'' where a music bot plays from a member-curated playlist, and all other members are automatically muted. 

Server creators can create different ``roles'' with custom names that grant users different permissions in the server, through which moderators gain their permissions as well. This role system allows for a hierarchy of moderation with lower-level moderators having less permissions, and higher-level ones having more. Depending on permissions granted to a given role, moderators can mute people, deafen people, or remove people from voice channels. Some moderators can also ban people from their servers, who will not be able to rejoin unless they are ``unbanned.''

While the forms of punishment provided by Discord are permanent by default, third-party applications called ``bots'' can be used to augment moderation by adding timers, making these actions temporary. Bots like MEE6\footnote{https://mee6.xyz/}, Dyno\footnote{https://dyno.gg/}, and Tatsumaki\footnote{https://tatsumaki.xyz/} are well-regarded and widely used by over a million servers to automate existing Discord features such as sending welcome messages and assigning roles. Besides improving existing moderator tools, many bots also provide additional functionalities for moderators, such as issuing people warnings that are permanently recorded in a moderator-only channel, and automatically removing content in text channels based on keywords or regular expressions. However, to the best of our knowledge, there are currently no bots with voice moderation capabilities. 

\section{Method}
To understand moderators' experiences in moderating voice-based communities, we conducted in-depth, semi-structured interviews with moderators of Discord servers. Participants were recruited as part of a larger collaborative, IRB-approved project to investigate moderation in online communities. For this study we analyzed 25 interviews with moderators who identified as having experience in moderating Discord voice channels. We recruited participants by reaching out to moderators of open Discord servers. 
We also asked them to send the call for participation to other moderators, resulting in a snowball sample. The first two authors conducted the interviews.
The 25 participants came from 16 different Discord servers, with between 1 and 3 participants from each server. While the majority of the servers that we examined are large ones with more than one thousand members and may not be representative of smaller groups, we believe this over-representation is reasonable as formal moderation is less-needed in smaller communities \cite{seering_moderator_2019,kiene_technological_2019}.  Our moderator participants provided us with a diversity of perspectives both across and within communities. Each participant was compensated US \$20 for their time.

Interviews ranged in length from 42 to 97 minutes, all of which were conducted over Discord voice chat. Participants' ages ranged from 18 to 39 ($M=24$, $SD=5.43$). Details about the participants, including age, gender, country of residence, and type and member count of the servers they moderate are presented in Table \ref{tab:demographics}.

\begin{table}
\begin{center}
\caption{Participant details. Numbers of server members are as of March 8, 2019.}
\label{tab:demographics}
\renewcommand{\arraystretch}{1.2}
\begin{tabular}{cccccc}
\textbf{Participant ID} & \textbf{Age} & \textbf{Gender} & \textbf{Country} & \textbf{Server Type} & \textbf{\# Members} \\
\midrule

\rowcolor{lavenderblue}
P01 & 18 & M & Croatia     & Social Chatroom & 164,257 \\

P02 & 19 & M & US          & \multirow{2}{*}{Streamer} & \multirow{2}{*}{117,742} \\
P03 & 19 & M & Russia      & &                    \\ 

\rowcolor{lavenderblue}
P04 & 21 & M & US          & Tech Support    & 233,762 \\

P05 & 21 & M & India      \\
P06 & 20 & M & US & \multirow{-2}{*}{Anime} & \multirow{-2}{*}{130,924}                 \\

\rowcolor{lavenderblue}
P07 & 18 & M & UK          &  &  \\
\rowcolor{lavenderblue}
P08 & 20 & F & Malaysia & \multirow{-2}{*}{Social Chatroom} & \multirow{-2}{*}{57,319}\\

P09 & 22 & M & US          & NSFW            & 23,186 \\

\rowcolor{lavenderblue}
P10 & 23 & M & UK  & &        \\
\rowcolor{lavenderblue}
P11 & 39 & M & UK   & \multirow{-2}{*}{Gaming} & \multirow{-2}{*}{29,165}       \\

P12 & 23 & F & US          & Fandom          & 150       \\

\rowcolor{lavenderblue}
P13 & 24 & M & Australia   & NSFW            & 55,239       \\

P14 & 19 & M & US          & Social Chatroom & 77,512\\

\rowcolor{lavenderblue}
P15 & 26 & M & US & &     \\
\rowcolor{lavenderblue}
P16 & 24 & M & US & \multirow{-2}{*}{Gaming} & \multirow{-2}{*}{55,251}   \\

P17 & 37 & M & US          & \multirow{2}{*}{Fiction Writing} & \multirow{2}{*}{1,137}  \\
P18 & 32 & F & US        \\

\rowcolor{lavenderblue}
P19 & 26 & F & US          & Gaming          & 3,246  \\

P20 & 24 & F & Netherlands & \multirow{3}{*}{Gaming} & \multirow{3}{*}{24,542} \\
P21 & 27 & M & US           \\
P22 & 22 & F & US          \\

\rowcolor{lavenderblue}
P23 & 23 & M & Netherlands & Gaming          & 171,608 \\

P24 & 24 & F & UK          & \multirow{2}{*}{Gaming} & \multirow{2}{*}{63,001} \\
P25 & 29 & M & US          \\

\bottomrule

\end{tabular}
\end{center}
\end{table}

During the interviews, we asked participants to tell us specific stories about moderating voice channels, with follow up questions about how they found out about rule breaking, what specific actions they took, and what impact the incident had on the moderation team as well as the community. We also asked them to consider hypothetical scenarios, such as what participants would do if the rule breakers tried to evade punishment. Participants detailed a variety of moderation experiences that ranged in scale and in complexity. We also asked participants about the challenges of moderating voice channels, the behind-the-scene deliberations of their moderator teams, and their feelings toward decisions they had made or situations they had encountered. Prior to analysis, all interviews were transcribed, anonymized, and assigned the participant IDs presented here. 

We performed a thematic analysis of the interview transcripts \cite{braun_using_2006}. The first author initially engaged in one round of independent coding, using an inductive open coding schema. All authors then discussed preliminary emerging code groups such as ``catch in the act,'' or ``enter the voice channel to confirm.'' Two more rounds of iterative coding helped us combined similar groups to create higher order categories such as ``moderation challenges.'' The first author used these categories to produce a set of descriptive theme memos \cite{saldana_coding_2009} that described each category with grounding in the interview data. All authors discussed the memos regularly to reveal the relationships between the categories and finally clarified the themes, which resulted in the three main findings we discuss below.

\section{Findings}

In describing the findings of this study, we start by characterizing new types of rules and new ways to break these rules in voice channels, then compare them to common types of rule violations in text communication. We then discuss the actions that moderators take to address these rule violations. Finally, we address the biggest challenge of rule enforcement in voice---acquiring evidence---by discussing moderators' strategies to gather evidence and how they often fail.

\subsection{Rules in Voice and How People Break Them}

Formal rules on Discord exist at the platform level in the form of Terms of Service and Community Guidelines, as well as at a community level in the form of custom rules set by the individual Discord servers. All the servers in our study had at least some explicit rules that were listed in specialized text channels, as well as implicit rules that were not written down but were nevertheless enforced by moderators. Though there were likely also emergent social norms in these communities, and rules that may have started out as norms, we spoke to moderators about the rules that they actively enforced, whether explicit or implicit, as opposed to norms enforced by the community itself. While there were many rules in the servers we examined, here we only focus on those with elements unique to voice.


\subsubsection{Explicit Rules.}

Servers that we discussed with moderators had different explicit rules that governed both text and voice channels, such as ``no advertising'' or ``English only,'' but all 16 of them had a rule against slurs and hate speech. We choose to take a deep dive on the rule of slurs and hate speech because it is the rule that most participants talked to us about, and presented challenges unique to voice. 

Slurs and hate speech can make a community an unwelcoming and unsafe space for its members, and therefore many communities have rules against them \cite{fiesler_reddit_2018}. Just like in many text-based communities, slurs and hate speech are explicitly prohibited in voice channels, and are a major problem that moderators have to face. All participants told us that racial and homophobic slurs existed widely in their servers, both text and voice channels. In P08's server, racial slurs in voice channels faced an even harsher punishment than in text channels:

\begin{itquote}
Racial slurs in the [text] chat and VC [voice chat] are different. If you say it in the [text] chat, you get a four-hour mute depending on the severity, and in the VC, you get an instant ban because it's more ... you know, saying it, rather than typing it, is much worse. (P08)
\end{itquote}

Racial slurs can be more offensive when spoken in smaller groups. Voice channels usually have 5 to 25 people participating at a time, which is much less than in text channels that typically have hundreds of active members. A potential consequence of the limited number of participants is that slurs in voice channels may feel more targeted and personal. 

While slurs were not allowed in any of the servers, how moderators determined the threshold for what counted as a slur varied. For example, P03 treated the slur ``n-{}-{}-er''and all of its intonations with a heavy hand:

\begin{itquote}
  Like if you were saying, the ``N'' and then ``word,'' then it's fine because it's not saying the slur itself. Any workaround ... is not allowed. ``N-{}-{}-a'' or ``n-{}-{}-a''---that's not allowed. Because you're still saying the slurs. Just rephrasing it. (P03)
\end{itquote}

Many moderators were cognizant of the different intonations a slur can have, and still decided to uniformly ban them---the only exception was the indirect reference ``n-word.'' At the same time, while P06 agreed that different varieties of racial slurs were still racial slurs, he also took context and intention into account in his moderation, and the intonations did matter in his decisions:

\begin{itquote}
I think there's a difference between saying ``What's good my n-{}-{}-a'' and ``I think all n-{}-{}-ers should go back to Africa.'' There's a pretty clear difference. So in that sense, you know, they're both racial slurs technically. And so by our rules ... the punishment would be the same for both. But you know, again, it's kind of like a case-by-case thing. (P06)
\end{itquote}

While the intonations can still be expressed in text, P06's quote suggests that voice introduces more nuances to moderation, and that what technically counts as racial slurs by explicit rules may still receive different treatment. In text-based communities, moderation of content and intonations can be automated by using a list of keywords or regular expressions (e.g., \cite{chancellor_thyghgapp:_2016}). But in voice communication where moderation cannot be as easily automated, and moderators have to hear everything for themselves, their personal judgments play a more important role. Having to moderate in a case-by-case manner also means more work for moderators.

\subsubsection{Implicit Rules.}

While slurs and hate speech were explicitly against the rules in the Discord servers we examined, we also heard about behaviors that were not written in the rules, but that moderators still discouraged or prohibited. While moderators technically had the option to detail these in their explicit rules, these behaviors were complex, nuanced, and difficult to articulate. Below, we focus on three main types of these behaviors that are unique to voice, and describe them as implicit rules in the servers: disruptive noise, music queue disruptions, and raids.

\paragraph{Disruptive Noise.}
Disruptive noise involves intentionally creating a loud or obnoxious sound in voice channels to irritate other people and disrupt conversations. According to many moderators, disruptive noise is a common rule violation in voice channels. One moderator, P14, said that their typical day involves ``muting at least one person'' in response to this kind of behavior. Disruptive noise points to several implicit rules that are not important in text-based communities, but stand out in voice. One of these rules is that one should not speak too loudly in a group conversation:

\begin{itquote}
I've had to step in because someone's told me ``Oh there's a kid literally screaming down in the \#underbelly'' ... So I'll hop in, and of course the kid will be screaming and he'll be muted. (P16)
\end{itquote}

The rule of not speaking too loudly shows a key difference between voice and text communication: voice is a limited-capacity communication channel. While typing in all caps does not affect other members' ability to type and be seen, speaking in high volume in a group voice channel takes up all capacity in the channel, effectively silencing others. Even though text spamming---posting lots of content repeatedly and quickly---also takes up channel capacity in a similar way, it is preventable by limiting the number of messages one can post in a given period of time (e.g., ``slow mode'' in Discord). Loud screaming, on the other hand, is not actively preventable on Discord unless a moderator steps in. Prior work has shown that people are already struggling with how to appropriately interject in group conversations \cite{isaacs_what_1994} with their assumed sequential nature \cite{resnick_grounding_1991}. The rule violation here is even more severe because it completely ignores turn-taking in conversations and forces the conversation to be about one person. 

However, volume by itself was not the golden rule of determining whether someone is creating noise. As P14 suggested, hardware conditions also had an impact on someone's speaking volume:

\begin{itquote}
[Disruptive noise is] typically anything that would be considered ``too loud.'' However, if someone has a sensitive mic, we typically let them know of this and make them aware how loud it is before taking action. (P14)
\end{itquote}

P14 further told us how he differentiated between intentional noise making and simply having a sensitive mic: 

\begin{itquote}
When someone joins a VC [voice channel] for the sole purpose of ``ear raping''\footnote{Though we intentionally excluded it from all but this quote, we would like to note that a number of our participants used the term ``ear raping'' to refer to creating disruptive noises. We hope that future work examines the use of the term in more depth, including the motivations behind its use and in what ways it may be harmful to individuals or communities on the site.}, they will typically do so immediately; someone who joins a VC to use it for its intended purpose will typically start off with a greeting or some sort of intro. I try to be aware that not everyone is capable of purchasing the best microphones, so I try to take that into consideration. (P14)
\end{itquote}

While it may be possible to develop an automated program that detects disruptive noise by decibel value, P14's quotes show that in addition to presented behaviors, intention also matters. If intention is important for determining rule violations, automated tools clearly have limitations. Furthermore, the fact that the way someone initially joins a voice channel is important suggests a challenge in moderation: A moderator has to be present when someone joins, which is not a scalable solution when voice channels are always open and a moderator is always required. This constant presence is also not a reasonable request for volunteer moderators who are only contributing during their free time.

In addition to the appropriateness of volume, disruptive noise as a type of rule violation also points to the appropriateness of content in group conversation:

\begin{itquote}
There is one time I had my Discord on speaker and I was just talking to a group of friends. ... [Some random people] joined and they started playing loud porn. So my brother was in the house ... and he heard the porn blasting out of my speakers and he was like, ``Yo dude, why are you listening to \emph{that} at full blast on speaker?'' (P06)
\end{itquote}

People tend to dislike unexpected auto-play audio on computers as it can cause physical discomfort and can be socially awkward in public \cite{ackerman_hanging_1997, chen_autoplay_2018}. The inappropriateness of the content here only exacerbated the situation. Furthermore, people not directly part of the voice channel were also affected, which demonstrates how moderation work can potentially have a negative impact on moderators' lives outside the servers they moderate. P06 told us that the experience was ``a little bit awkward,'' but it is likely that the same situation can happen in other contexts, such as in a workplace, with more serious consequences. 

We also heard a similar story of disruptive noise where the sound itself was not irritating, but rather the way the sound was played: 

\begin{itquote}
We'd have this weird phenomenon where there was a handful of people who would join voice for 5 seconds, leave for 2 minutes, then come back and join voice for 5 seconds, leave. So while you're playing all you would hear was ``boop'' ``do-doop'' [Discord notification sounds]---people just joining and leaving rapidly---and it was just so infuriating. (P10)
\end{itquote}

Just like in P06's example above, the unexpectedness also stands out in this example---people would not expect Discord features to be used in annoying ways. Furthermore, while it is possible to turn off these notification sounds, it is not feasible to do so in the middle of a game. There is also no way to ping a moderator without disengaging with the ongoing game, as in the case in P10's quote. This example also points to a difference between text and voice---in text channels, constant interruptions are accepted and somewhat expected. But in voice channels, the flow is much more important and even the slightest interruption can be disruptive.

\paragraph{Music Queue Disruption.}

Across our interviews, we heard stories of rule violations not only in conversational voice channels, but also in music queues---voice channels that automatically play from crowdsourced playlists. In music queues, members are automatically muted, but they can append new music to the shared playlist, and skip the music that is currently playing if the channel setting allows.

A rule that music queues shared with conversational voice channels was not to be too loud:

\begin{itquote}
Literally the most recent thing that happened. ... Someone put [something] in the music queue and it was for like two hours of just extremely loud music. (P04)
\end{itquote}

P04's quote shows that there was another type of rule violation other than being too loud: the music was also long. According to P04, checking whether people put ``hour-long shitposting tracks that nobody wants to listen to'' in the music queue was his everyday job. Because music queues are necessarily sequential, a turn-taking rule becomes important: people agree to occupy the channel completely for an appropriate, limited amount of time. Compared to disruptive noise, the problem with playing long music is not that it takes up the whole channel---it is in fact expected---but is that it essentially rids other members of their turns by taking up the channel for a long time.

Another form of rule violation that stops other people from participating is to skip their music: 

\begin{itquote}
I actually got mad at the user. ... What they were doing was they were constantly taking our jukebox and turning off other people's stuff. Like someone would play something and then they would skip it. (P21)
\end{itquote}

This example, together with the previous example, emphasizes the importance of turn-taking in music queues. While turn-taking is not a problem in text because of its threading structure, it becomes a major problem in music queues that resemble town hall meetings, where everyone gets an opportunity to hold the mic---playing long music is like someone who will not stop talking, and skipping music is like forcing the person with the mic to stop talking. 

\paragraph{Raids.}

In addition to rule violations by individuals, we also heard stories of organized rule violations that moderators called ``raids.'' Raids are organized voice channel disruptions that involve multiple users that can be human or bots. P13, for example, experienced a raid that was similar to P06's story, but on a larger scale:

\begin{itquote}
There was one a few months ago where they were spamming porn over their mics and they all had profile pictures of the same girl in a pornographic pose. And there were maybe like 15 of them in the same voice chat. (P02)
\end{itquote}

While an individual playing pornographic audio is irritating by itself, one can only expect 15 people doing so would cause even more discomfort. While raiding violates the similar rules in voice that we mentioned above, it is considerably more difficult for a moderator to manage. In the case of an individual, a moderator only needs to take action on that person, but in the case of a raid, a moderator needs to act on all the people involved. Across our interviews, we heard stories of raids that involved up to thousands of bot accounts, but moderators could only manage the accounts one by one---there is currently no way to manage multiple accounts all at once. This restriction means that not only do the moderators have to take on a significant amount of work managing raids, but also there is no way to prevent the other raiders from evading once they see one of them is punished.

Fortunately, while managing raids could be difficult, recognizing raids was relatively easy. As P02's quote suggested, the raiders all had similar profile pictures, which often became a clear signal for raid detection:

\begin{itquote}
[W]hen I see multiple people with the same or similar names rapidly join a VC, that's a warning sign for me. (P14)
\end{itquote}

Moderators told us that when they saw people with similar profile pictures or names, they would join the voice channel to confirm if they were part of a raid. However, we also heard from some moderators that they saw these cues as definitive signals of raids, and punished these people without confirming. Moderators told us that there were no unfair punishments because they believed if there had been any, these people would have appealed. This moderation approach can be potentially problematic in cases where members use similar usernames as part of a community in-joke, which we also heard in the interviews. While moderators' tolerance of false positives---incorrectly identifying someone as a raider---may be a reasonable attempt to reduce their workload, it also suggests that they could have unknowingly driven these members away for good.

\subsection{Moderation Practices}

To punish the rule breakers, moderators used tools provided by Discord, including muting, deafening, and banning for more serious offenses, which third-party bots enhanced by setting a timer on them, making these punishment temporary. While we found that the correspondence between rule violations and types of punishment was highly contextual to individual servers, there were still some common moderation approaches that the moderators took. Specifically, moderators tried to warn first before taking more severe actions, but sometimes they also took these actions merely based on hearsay or first impressions. Even though these signals were unreliable, moderators did not come up with specific rules that described acceptable and unacceptable behaviors, but developed catch-all rules that dictated their authorities instead. 

\subsubsection{Warn Before Punishment.}
Across our interviews, a common approach to moderate voice channels that we heard from 23 of 25 moderators was to warn first before giving out real punishment:

\begin{itquote}
We would unmute our mics, start talking casually to people and then we would just figure out if they are breaking the rules or not, warn them verbally in the voice channel because we do roll with things in voice channel. ... We'd always do verbal warning before we do any other actions. (P07)
\end{itquote}

P07's quote points to a key difference between voice moderation and text moderation. In the case of text, moderators would not have to ``figure out'' if someone broke the rules---all the conversations are recorded and visible to moderators. However, because Discord does not record voice channels, there is no way for moderators to unambiguously determine rule violations. ``Rolling with things'' allowed moderators to learn the contexts of alleged rule violations while not upsetting community members, and giving warning first let moderators still enforce rules to some extent but not have to risk wrongly punishing someone.

According to the moderators, warning was not only a way to conservatively enforce rules, but also a lesson for the community:

\begin{itquote}
We have this weird thing where our moderation is kind of public in that, if we warn someone not to do that, we try to always do that publicly so that other people can say, ``Hey, okay, that's an example of what not to do.'' (P19)
\end{itquote}

The public moderation that P19 mentioned suggests community members learn rules through examples, which prior research shows can be effective in discouraging undesired behaviors \cite{seering_shaping_2017}. While example-setting certainly has been successful in moderating text-based communities, explicitly showing what the rules are may be more important in voice-based communities because the rules might be ambiguous or unfamiliar, especially to newcomers who are more familiar with text-based communities. 

\subsubsection{Punishment Based on Hearsay and First Impressions.}

While in most cases moderators tried to give warnings first, they sometimes also punished rule breakers directly. For example, P11 told us about a time the moderation team banned a member only based on a member report, without extra deliberation:

\begin{itquote}
Someone complained that a user was harassing them in voice chat and just shouting profanities and racism down the mic, so we just banned the user and we didn't hear anything, and they didn't appeal it. ... We kind of took it at face value. It's very hard to get evidence from a voice chat unless you're recording. (P11)
\end{itquote}

Here, P11's moderation team assumed the punishment was fair only because the person punished did not push back, which may be an acceptable solution without adequate evidence. However, punishment based on hearsay does risk wrongly punishing someone in the case of false reporting. In such a case, the person punished by mistake possibly would not have appealed anyway when they were frustrated by undeserved punishment and left the community \cite{lampe_follow_2005}.

Moderators sometimes also took actions based on their own understanding of the person involved. P21, for example, told us that ``first impressions matter, especially in voice chat,'' to the extent that they justified the most severe form of punishment:

\begin{itquote}
So if a user has just joined ... and immediately shows this kind of behavior, such as the derogatory terms, trolling, sexist behavior ... they'll just get banned. We won't warn them. We'll ban you because you've made it quite apparent that you had no interest in reading the rules and you're not all that great of a person, to say the least. (P21)
\end{itquote}

P21's quote shows a stark contrast with the ``warning first'' approach of moderation, which he also took in his community. However, the importance of first impressions in voice suggests that a person's intention and character may be more salient than in text, and they can be used in moderation decisions. 

\subsubsection{Catch-all Rules.}

With moderators sometimes having to use unreliable cues like first impressions in moderation, one question arises: Couldn't they develop rules that unambiguously describe what behaviors were acceptable or not? Moderators gave their answers to this question:

\begin{itquote}
There are times where you can't catch everything with a rule set. So I think that's more the reason why we don't have an on paper set of rules because we are such a large server. We would have to have a million different rules to help cover everything. (P04)
\end{itquote}

Across the Discord servers we examined, only two had guidelines about acceptable behaviors specific to voice channels. P04's quote shows that the lack of guidelines comes from the practical standpoint that it is simply impossible to list every single type of behavior, which suggests that the variety of (mis)behaviors possible in voice is much more diverse than in text. However, this is not only a problem of quantity---the level of nuance involved in the rule violations themselves shows that the line between acceptable and unacceptable behaviors is also difficult to articulate.

The solution that servers used to solve this problem, including the two that had guidelines in voice channels, was to create catch-all rules that were different varieties of the same core idea: Moderators have the utmost authority. 

\begin{itquote}
[If someone does something unacceptable] then they are well within their rights to turn around and say, ``Well that isn't in the rules.'' And then it's just a nice get-out clause for the moderator to be able to turn around and say, ``Well look, whatever a moderator says goes.'' (P11)
\end{itquote}

This umbrella rule did not only free the moderators from having to create new rules, but also reduced the potential work of arguing with members. As moderators had greater authority, they also took greater responsibility in making the right decisions. But as we heard from the moderators, the lack of evidence in voice became an obstacle. In the next section, we describe the ways moderator gathered evidence of rule violations and their unique challenges.

\subsection{Acquiring Evidence}
Moderators told us about their various practices and actions upon rule violations, but a prerequisite of any  action is that moderators had to make sure that a rule was violated. Acquiring such evidence in ephemeral voice channels, however, was a major challenge:

\begin{itquote}
Voice channels just basically can't be moderated. ... The thing is in voice, there's no record of it. So unless you actually heard it yourself, there's no way to know if they really said it. (P19)
\end{itquote}

Gathering evidence was such a challenge that moderating voice channels was almost impossible to P19, and this difficulty points directly to a fundamental difference between voice and text. In text, it is common knowledge that every conversation is automatically recorded. Moderators would not even have to consider acquiring evidence because everything is persistent. Even for ephemeral text chat applications like Yik Yak, where community moderation mechanisms such as upvoting and downvoting are possible \cite{schlesinger_situated_2017}, the text would still have to be persistent for a short period of time. However, this common assumption about text-based moderation breaks down completely in voice because of the ephemerality of real-time voice conversation. 

Moderators came up with different strategies and workarounds to tackle the problem of obtaining evidence in an ephemeral environment. In the rest of this section, we describe three main types of strategies that we heard from moderators: (1) Entering voice channels when the rule violation was happening; (2) asking witnesses to confirm someone was breaking the rules; and (3) recording voice channels.

\subsubsection{Entering Voice Channels to Confirm Rule Breaking.}

As P19 said, the most reliable way for moderators to acquire evidence was to hear it for themselves. However, constant presence in voice channels is not a reasonable request for volunteer moderators. Therefore, moderators took one step back---entering voice channels when they realized someone might be breaking the rules:

\begin{itquote}
When we do get a report, then if someone is available and it's going on right then and there, then we'll hop on and listen and see what's going on and see if we can determine any rule violations right there off the bat. (P25)
\end{itquote}

While entering voice channels may be a solution, P25's quote does point out a important requirement: a moderator has to be online at the time of rule violation, and the rule violation has to be still ongoing when the moderator joins. This requirement is difficult to fulfill, considering the time a member would take to report to a moderator, the time the moderator takes to see the report, and the time the moderator takes to join the voice channel. This requirement also means that, for any violations that are instant and do not extend over a period of time, moderating with this method is nearly impossible.

Even if a moderator was present at the right time, it was still not guaranteed that the moderator could identify the rule breaker:

\begin{itquote}
There's also the problem of telling who's talking, because ... if everyone's talking at once and some big fight is happening, it's kind of hard to tell who's talking. (P21)
\end{itquote}

Discord has no platform limit on the number of people a voice channel can have, and many moderators had moderated large voice channels with as many as 25 people.  Voice channels, unlike text channels with threading structures, is inherently limited in information capacity. Therefore, it can be difficult to tell who is saying what when many people are talking simultaneously. Furthermore, there was also another problem with multiple people in the same voice channel---there was no way to stop people from evading:

\begin{itquote}
When they start seeing people get banned, they leave the voice chat so they don't get banned themselves and then it makes it harder for us to find them. (P02)
\end{itquote}

Tracking down rule breakers, according to P20, was often a ``wild goose chase, only on steroids.'' The need to track down rule breakers speaks to the importance of not only acquiring evidence, but also the information contained in the evidence, which is also different between in text and in voice. In text, the text itself contains evidence in the form of rule-breaking content (e.g., racial slurs), but also the metadata---who the poster was, when it was posted, etc. When a moderator need to punish a person, this information connects the problematic content to the person, allowing the moderator to act without the person and the content being present at the same time. In real-time voice where content is ephemeral, however, this metadata does not exist. Even though the moderator has the evidence---they heard the content---there is no persistent record that associates the content to the rule breaker. Furthermore, identities in voice channels are also ephemeral: once the person leaves the voice channel, their profile picture disappears, and the information that connects the voice to the person no longer exists. Without this metadata, in text the moderator can at least delete the content, but in voice, the content is ephemeral to start with, and the moderator can neither moderate the content nor punish the person.

\subsubsection{Relying on Witness Reports.}

With joining voice channels being ineffective in many cases, some moderators turned to witnesses' testimony as evidence:

\begin{itquote}
If the violation has already passed and people have split up, then we'll get a list of the users that were in that voice channel at the time and we'll send a message to them and just ask them, ``Hey, you heard that there was something going on with this voice channel, can you give us more information about this?'' (P25)
\end{itquote}

Moderators hoped to use witnesses' memories as a way to overcome the ephemerality of voice, but note that P25 here contacted a list of members instead of a single member. Without concrete evidence, the quality of witness reports became important to separate legitimate reports from hearsays and rumors, and one proxy for quality was the number of witnesses:

\begin{itquote}
We need witnesses like at least maybe two other people ... that were there, that can confirm like, ``Yes, this person said that.'' (P02)
\end{itquote}

P02's quote suggests a simple rule of thumb: the more witnesses there were, the more credible the report was. While moderators spoke of this as a generally successful approach, there were nevertheless problems with it. First, multiple witnesses would work only if their stories were consistent, but when they were not, moderators were likely to fall into the rabbit hole of seeking more witnesses and more stories. This could lead to the moderation workload no longer being comparable to the severity of the violation anymore. Second, even if the stories were consistent, moderators had no way to make sure the stories were true:

\begin{itquote}
You just have to be careful with the report, because some people can do false reporting, to do whatever to the other user. (P24)
\end{itquote}

P24's quote reveals an interesting possible scenario, where one member falsely accuses another member due to some personal vendetta. Requiring multiple witnesses may be able to mitigate this problem between dyads, but could facilitate a planned brigade against a community member, and in this case the member could not even appeal their case, again, due to the lack of concrete evidence. To mitigate the problem of false reporting, some moderators came up with a workaround---sending ``spies'' into the voice channel:

\begin{itquote}
We have these accounts that don't have the staff role, you know, you just send those accounts and if someone catches them they're able to get warned. ... You're basically sending someone who is not a staff to catch that guy. (P05)
\end{itquote}

Three moderators told us that they asked trusted members to be ``spies,'' or sent in undercover moderators into voice channels. This practice shows the complexity of moderators' evidence acquisition strategy, and also suggests a difference in the expectation of privacy between text and voice. In text, people assume that everything they write will be accessible by the moderators, but real-time voice chat breaks down this notion and suggests an increased amount of privacy---people cannot hear what someone is saying unless they decide to join the same voice channel and be seen. When there are accounts that do not truthfully represent their identities, people in the voice channel no longer know (1) who they are talking to, or (2) who can hear their conversation. One may think that people should not break the rules anyway and they will not be impacted if they do not break the rules, but this \emph{de facto} surveillance system can create a chilling effect that discourages members from using voice overall.

\subsubsection{Recording.}

Recording voice channels is probably the most straightforward way to address the problem of evidence by making ephermeral content persistent, and allows moderators to discover rule-breaking voice in a similar way in text. However, instead of going into the voice channels and recording directly, moderators took more concealed approaches. P21, for example, used a method that he called ``incognito recording'':

\begin{itquote}
There were six users in voice chat ... harassing a female player ... so I was alerted and I joined and they immediately clammed up. I left and ... I had a user [who was not a moderator] join the voice chat and I recorded through that user to listen to the entire conversation. (P21)
\end{itquote}

P21 used a combination of voice recording and ``spy'' users, because the rule violators would immediately stop when a moderator joins. This example further breaks down the notion of privacy in voice channel described in the last section: people no longer know whether their conversation is on the record, on a platform where they expect ephemerality. Prior research showed people had strong negative reactions toward instituting permanence in a previously unarchived space \cite{hudson_go_2004}, and we can only speculate that doing so secretly will even further discourage people from participating. 

While recording might be a guaranteed way for moderators to get evidence, moderators told us recording only voice was not enough:

\begin{itquote}
We only really take MP4 files at the moment ... and what it will do is it will come up with a little snippet of what the voice channel actually looks like with all the people in it. And because whenever someone talks in voice channel, it has a green circle around their avatar. What we do is we fathom out who's talking at that moment saying whatever they're saying and then the action it needs. (P07)
\end{itquote}

For open servers, it is impossible for moderators to connect voices to members' identities---moderators cannot remember everyone's voice and cannot possibly have heard everyone's voice in a server with tens of thousands of people, not to mention there are new people joining all the time. Screen recordings, according to P07, mitigated this problem because they indicated who was talking at the time (in this case green circle around speakers' profile picture), and essentially achieved the effect as if moderators were there hearing the conversation for themselves. However, one problem we identified with moderators entering the voice channels---they could not tell who was breaking the rules when multiple people were talking at the same time---still persists with this solution. Furthermore, screen recording also introduces more technical overhead because the resulted file size would be considerably larger than textual recording or even voice recording. The problem of screen recording resonates with Grimmelmann's discussion of the need to contain the cost of moderation within acceptable levels \cite{grimmelmann_virtues_2015}.

Finally, one moderator raised a problem with recording that extended to the legal space:

\begin{itquote}
Recently we had like one example of, somebody reporting somebody for recording them in the voice channel, which is like a can of worms because they were like, according to Turkish regulation, or whatever where this user is from, it's illegal to record somebody without two party consent. (P23)
\end{itquote}

Laws regarding recording conversations are not consistent around the world. In eleven states in the United States, as well as in some other countries such as Germany and Turkey, two-party consent--- that is, all parties of the conversation must consent to the recording---is required \cite{noauthor_telephone_2019}. Sending the recording to other people is also against the law in countries like Denmark, which could potentially deem the behavior of the member who recorded for moderators illegal. However, currently there is no consent process for voice channel recording in either Discord's Terms of Service (other than a term that prohibits illegal activities), or the rules of the servers that we examined.

Overall, our findings revealed new kinds of behavior and norms that necessitated new rules in the context of voice communication, as well as new ways for people to break them. Moderators developed various practices to adapt to these changes, but they struggled to acquire evidence of rule breaking. While they developed tactics and workarounds to gather evidence, none of them were perfect, and all of them introduce problems that ranged from punishing an innocent community member to potentially breaking the law. These problems are unique to moderating voice, and challenge existing assumptions about moderation practices that are common in text. In the discussion section, we first detail the difference between voice and text and how they impact community moderation. Then, we provide design implications for platforms and individual communities, as well as lessons for emerging communities that use new types of social technology.

\section{Discussion}
The stories we have heard revealed many unique moderation problems in voice that moderators had difficulty solving. Moderators were often unable to prove someone broke a rule, and when they did, they could only moderate in a post-hoc manner. In this section, we first discuss the fundamental differences between moderating voice and text through the lens of Grimmelmann's discussion of moderation in online communities \cite{grimmelmann_virtues_2015}. Building on that analysis, we highlight implications for design and spaces where platform designers and administrators can improve the status quo.

\subsection{The Unique Challenges of Voice}

Participants' stories about moderation in voice channels challenge many understandings and assumptions about moderation based on text-only communities. In his analysis of text-based communities, Grimmelmann lists four moderation techniques, or ``the basic actions moderators can take'': excluding, pricing, organizing, and norm-setting. All of these techniques are available in text-based communities, but real-time voice communication rids moderators of many of them.

In open text-only communities, organizing seems to be the most common moderation approach based on prior work (e.g., \cite{kiene_surviving_2016, seering_moderator_2019}): deleting posts, editing the content of posts, annotating posts (such as adding tags and flair), etc. However, these basic, commonly-used abilities completely cease to exist in real-time, ephemeral voice chat: there is currently no way to delete, edit, or annotate someone's voice as they speak. In open communities where pricing is not an option, moderators of voice-based communities are left with only two options: excluding rule breakers from the community, or setting desired norms. However, neither of these is without its problems.

In text-only communities, strategies that exclude rule breakers, such as muting or banning them, are typically a last resort for severe cases since moderators are able to moderate the problematic content itself. As long as the case can be handled in terms of only content, actions on the person who posted the content are usually unnecessary. On the other hand, in voice channels, banning or muting someone is the only way to prevent problematic content from appearing, because that content is entirely dependent on the person's presence. In other words, banning content is equivalent to banning a person's presence in voice, either permanently or temporarily.

Norm-setting is also challenging in voice-based communities. According to Grimmelmann \cite{grimmelmann_virtues_2015}, moderators can set norms directly by making rules, or indirectly by showing examples. However, our findings suggest that moderators avoid direct norm-setting due to the greater variety and complexity of ways to break rules in voice. Compared to text, it is less feasible for moderators to create a separate rule for every possible violation, not to mention some of which might be too nuanced for rules to articulate. While our findings show that moderators do engage in public example setting as a way to indirectly influence norms, it can take a long time for norms to emerge and moderators still have to take the more extreme action of excluding in order to meet their short term need to remove problematic content. 

In addition to rendering these basic actions unusable, voice also restricts moderators in terms of the tools they can use. For example, moderators cannot auto-moderate voice because there is currently no equivalent to keyword-filtering. Instead, moderators can only react because there is no way to preemptively prevent rule violations from happening in voice at the level of infrastructure. Compared to text, in which moderators have many tools at their disposal, in voice moderators are fighting with their bare hands, against more and harder problems. As articulated by Clark and Brennan \cite{resnick_grounding_1991}, the medium used can dramatically change the availability, cost, and effort required for communication techniques; our analysis reveals that these challenges due to change in medium apply to not just grounding, but also moderating communication.

Beyond the challenges of executing moderation when they know that a rule has been broken, there are additional challenges for moderators to even determine \emph{whether} a rule has been broken. Our participants emphasized a critical aspect of their practice that does not appear in prior work focusing on moderation tools and strategies (e.g.,\cite{grimmelmann_virtues_2015, kou_regulating_2013}): the need for evidence. In text, there possibly is no such need---because until a moderator deletes it, text will always be there. On the other hand, moderators told us that evidence was a major problem in voice channels, due to the ephemerality of voice. While moderators developed strategies to address this issue, such as entering voice channels when receiving a report, or recording, these strategies were unreliable at best, and at worst, they risked breaking the law. 

\subsection{Designing for Voice Moderation}

These challenges to voice moderation that our findings reveal point to several implications for the design of moderation tools. First, many rule violations that moderators identified in voice channels may be preventable using automated systems. Moderators told us that slurs and hate speech are common in voice channels. It is tempting to consider the use of automated systems to detect this type of speech, as many moderators already do in text-based communities \cite{seering_moderator_2019}. However, our findings also show that the intonation can be nuanced, and moderators take context into account and make moderation decisions on a case-by-case basis---which is challenging for an automated system to do. Furthermore, these systems may unfairly punish those who do not speak English in the same way that these voice recognition systems were trained \cite{harwell_accent_2018, paul_voice_2017}. As Grimmelmann \cite{grimmelmann_virtues_2015} pointed out, any moderation action can nudge the community norms in an unpredictable direction, and a blanket ban of what automated systems deem inappropriate may do more harm than good for a community. This caveat applies to any moderated online community, but the potential harm may be more pronounced in communities in which the underlying rules and norms are evolving or unclear, as is the case in many Discord voice channels.

To prevent disruptive noise, for example, platform or system designers can design systems that detect volumes that may be uncomfortable for human. An intuitive implementation of this system would be to automatically mute accounts that are too loud, but our findings suggest that loudness can be a result of misconfigured hardware. Therefore, while temporary muting would still be necessary, the system may also want to prompt loud accounts to check their hardware settings. Similarly, to mitigate music queue disruptions caused by lengthy audio, a system could alert moderators of audio that exceeds a certain length threshold, and let them determine whether it is a rule violation. Both of these design recommendations aim to prevent rule violation preemptively, rather than reactively.

To address the major need for moderators to acquire evidence, platforms may want to incorporate recording functionality within moderator tools. While there are already third-party applications that are able to record voice channels on Discord, such as MEE6, it may still be better for platforms to have control over such features. Third party applications may pose privacy risks to users, but if implemented within the platform, Discord could both take measures to protect the recorded data, and make sure that users are informed. Furthermore, without access to the platform directly, third party applications typically only record audio, which, as our findings suggest, is not sufficient when moderators need to connect distinct voices to user accounts. Therefore, platforms may also need to generate video files, or audio files with metadata, that show who is speaking (or not) at any time.

However, we also recognize that the design recommendations above largely require some type of automated system to listen in the voice channels at all times, which may raise privacy concerns among community members. This issue, together with our prior discussion of ``incognito recording,'' steps into the legal and policy realm of recording conversations. One-party consent recording---recording by a participant in the conversation without other parties' consent---is against the law in eleven states in the U.S. and some other countries \cite{noauthor_telephone_2019}. Discord's Terms of Service does not have rules about recording, nor any other behavior specific to voice. However, it does prohibit users from ``engaging in conduct that is fraudulent or illegal,'' and require users ``to comply with all local rules and laws regarding your use of the Service, including as it concerns online conduct and acceptable content,'' recognizing that Discord is an international platform. These blanket statements mean that the work of deliberating different regulations around the world is offloaded to the individual server moderators---something Discord explicitly states in its Community Guidelines: 

\begin{itquote}
We do not actively monitor and aren't responsible for any activity or content that is posted; however, we expect server owners, mod[erator]s, and admins to uphold their servers to these guidelines and we may intervene if community guidelines are not upheld.
\end{itquote}

However, it is not reasonable to assume that all moderators would know all the regulations in the world, particularly since people often have incorrect interpretations of both the law and Terms of Service provisions \cite{fiesler_reality_2016}. One of our participants stated confidently that recording is definitely not against Discord's Terms of Service---though given the complexities of the broader laws that the document nods to, this may or may not be true. Despite these complexities, our findings show that recording could be a desirable solution to the major problem of gathering evidence. Therefore, to prevent volunteer community moderators from bearing legal consequences unknowingly, platforms like Discord could either explicitly acquire consent at the platform level (e.g., in Terms of Service, or through a popup dialog when a user joins a voice channel for the first time), or advise individual community moderators to explicitly acquire consent within their communities if they wish to record voice channels.

\subsection{Beyond Voice}

Voice as a technology has been around for decades, but the need to moderate it in the context of an online community is a more recent phenomenon. While this study only focuses on moderating voice, online communities are likely to continue to develop beyond text and voice. Though specific issues may not generalize beyond voice, the types of problems we identify could appear in other new social technologies where moderation may be necessary. For example, we are already seeing emerging social VR communities---for example, VRChat, where people interact with virtual avatars and communicate using voice. VRChat shares many characteristics with Discord voice channels: interactions happen in real time and are not recorded. However, the virtual physical presence of VR adds another complication to moderation---for example, there are already reports of users who sexually harass other users physically \cite{noauthor_vrchat_nodate, noauthor_heres_nodate}. Our study identifies potential problems and sheds light on design opportunities for similar online spaces where interactions are real-time and ephemeral. 

It is important to recognize that moderation is not one-size-fits-all. Our findings point out that existing moderation strategies in one type of technology can break down completely in another. Therefore, while it may be easy and intuitive to import existing rules to a new community, we argue that designers and moderators should not ignore the technological infrastructure of the community when doing so, and carefully consider the limitations imposed by it.

It can be difficult to predict how people will abuse new technology, nor how rules or enforcement practices may need to change to prevent such abuse. Therefore, it is important that moderators are willing to change rules or make new rules as their communities adopt new technology. While our findings show that many of the rules were implicit in the context of new technology, we recommend moderators frequently reflect on their practices and consider whether implicit rules should be made explicit, so that new and old members can easily learn the rules. 

Finally, as moderators in different communities use different strategies to moderate, it is likely that moderators in one community already have solutions to problems in another community. While we did not hear about any cross-community collaboration in our interviews, reaching out to or forming groups with moderators from other communities can be a good way to collectively interrogate and mitigate problems brought by new technology.

\section{Conclusion}
While moderation is commonplace in online communities where moderators can remove, edit, and annotate content, change in technology can break down our assumptions and challenge our understandings of moderation by making these actions invalid. This study, through the lens of voice-based communities, examined new rules and new ways to break them through voice communication, and moderators' challenges without the tools that they used to take for granted. We found that not only did the moderators have a hard time enforcing the rules, proving that a rule violation even happened was also a challenge for them. While recording may be a desirable solution, our findings pointed out that voice-only recording was not enough because moderators were not able to link members' unique voice to their identities. Furthermore, the vague platform policies on recording may subject the volunteer moderators to legal ramifications. We argue that platform designers and administrators should carefully consider the impact of technological infrastructure on communities that they host, and develop technology and policies that provide moderators ample support and protect them from unexpected harm caused by the technology.

\begin{acks}
We would like to thank the reviewers for their significant time and care in improving this paper. We would also like to thank Jordan Wirfs-Brock, Brianna Dym, Benjamin Mako Hill, Morgan Klaus Scheuerman, Joseph Seering, and Kenny Shores. The first author would like to thank Arcadia Zhang, as always. Finally, we would like to thank our moderator participants who contributed their time to make their communities and Discord better, and ultimately make this project happen.
\end{acks}
\bibliographystyle{ACM-Reference-Format}
\bibliography{discord}

\end{document}